\documentclass[useAMS,usenatbib]{mn2e}

\usepackage[T1]{fontenc}
\usepackage[latin1]{inputenc}
\usepackage{times}
\usepackage{epsfig}
\usepackage{amsmath}
\usepackage{amssymb}
\usepackage{graphicx,subfigure,xspace}
\usepackage{url}

\newcommand{\G}[1]{\ensuremath{\Gamma(#1)}\xspace}
\newcommand{\C}[1]{\ensuremath{C_{#1}}\xspace}
\newcommand{\cl}{\C{\ell}}

\newcommand{\Pmat}{\ensuremath{\mathrm{\mathbf{P(0)}}}\xspace}

\title[$C_{\ell}$ and $\Gamma_m$ connection...]{About the connection between the 
$\bmath{C_{\ell}}$ power spectrum of the Cosmic Microwave Background and 
the $\bmath{\Gamma_{m}}$  Fourier spectrum of rings on the sky}

\author[R. Ansari et al.]{R. Ansari$^1$, S. Bargot$^1$, A. Bourrachot$^1$, F. Couchot$^1$, 
J. Ha\"{\i}ssinski$^1$,
S. Henrot-Versill\'e$^1$, 
\newauthor 
G. Le Meur$^1$,
 O. Perdereau$^1$, M. Piat$^2$,
 S. Plaszczynski$^1$ and F. Touze$^1$\\
$^1$ Laboratoire de l'Acc\'el\'erateur Lin\'eaire, IN2P3-CNRS and Universit\'e
de Paris-Sud, BP 34, 91898 Orsay Cedex, France \\
$^2$ Institut d'Astrophysique Spatiale, INSU-CNRS and Universit\'e 
de Paris-Sud,  91405 Orsay Cedex, France
}

\begin{document}
\date{original form 2003 January 16} 
\maketitle 

\begin{abstract}
In this article we present and study a scaling law of the 
$m\Gamma_m$ CMB Fourier spectrum on rings which allows us 
(i) to combine spectra corresponding to
different colatitude angles (e.g. several detectors at the focal
plane of a telescope),
and (ii) to recover the $C_\ell$ power spectrum once 
the $\Gamma_m$ coefficients have been measured.
This recovery is performed numerically below the 1\% level for 
colatitudes $\Theta> 80^\circ$ degrees. 
In addition, taking advantage of the smoothness of the $C_\ell$ and of the
$\Gamma_m$, we provide analytical expressions which allow
the recovery of one of the spectra at the 1\% level, the other one being known. 
\end{abstract}


\section{Fourier analysis
of circles on the sky versus spherical harmonics expansion}

Cosmological Microwave Background (CMB) exploration has recently made 
great progress thanks to balloon
born experiments (\citealt{boo}, \citealt{max} and 
\citealt{arc}) and ground-based interferometers 
(\citealt{cbi}, \citealt{das}, \citealt{vsa}).  
MAP\footnote{Map home page:\\ \url{http://map.gsfc.nasa.gov/}} 
whose first results will be available at the beginning of
 2003 and the forthcoming
Planck satellite\footnote{Planck home page:\\\url{http://astro.estec.esa.nl/SA-general/Projects/Planck/}} 
whose launch is scheduled for the beginning of 2007
will scan the entire sky with resolutions of 20 and 5 minutes of
arc respectively. These CMB observation programs yield to a large amount of data whose reduction is usually performed
through a map-making process and then by expanding the temperature 
inhomogeneities
on the spherical harmonics basis:
\begin{eqnarray}
\label{eq:one}
\frac{\Delta T(\vec{n})}{T}=\sum_{\ell}\sum_{m=-\ell}^{\ell}
a_{\ell m} Y_{\ell m}(\vec{n})\, .
\end{eqnarray}
The outcome of the measurements is given in the form of
the angular power spectrum $C_{\ell} \equiv <| a_{\ell m}|^2>$.
The set of $C_{\ell}$ coefficients completely characterizes the CMB 
anisotropies in the case of uncorrelated Gaussian inhomogeneities (\citealt{2002ARA&A..40..171H},
\citealt{1987MNRAS.226..655B}). 

Several of the current or planned CMB experiments (ARCHEOPS, 
Map, Planck) perform 
or will perform circular scans
on the sky.
Carrying out a one-dimensional analysis 
of the CMB inhomogeneities on rings provides
a valuable alternative to characterize its statistical 
properties (\citealt{del}). 
A ring-based analysis looks promising {\it e.g.} for the
Planck experiment where repeated ($\sim 60$ times) scans
of large circles with a colatitude angle $\Theta \sim 85^{\circ}$
are being planned. This approach
differs in several ways from the one
based on spherical harmonics. In particular it does not require the
construction of sky maps and some systematic effects could 
be easier to treat in the
time domain 
rather than in the two-dimensional ($\Theta,\varphi$) space 
($1/f$ noise for instance)
since the map-making procedure involves a complex projection 
onto this space.\

For a circle of colatitude $\Theta$, one writes
\begin{eqnarray}
\label{eq:prem}
\frac{\Delta T(\Theta,\varphi)}{T}=\sum_{m=-\infty}^{+\infty}
\alpha_{ m}(\Theta)e^{im\varphi}\,,
\end{eqnarray}
and the $\Gamma_{m}$ Fourier spectrum is defined by 
\begin{eqnarray}
<\alpha_{m}\alpha_{m'}
^{*}> = \Gamma_{m}(\Theta)\delta_{mm'}\,.
\end{eqnarray}
\noindent These $\Gamma_{m}$ coefficients are thus specific to a particular
colatitude angle $\Theta$. We propose below a simple way 
of combining sets of such coefficients corresponding to different
$\Theta$ values (i.e. different detectors).

\begin{figure}
\epsfig{file=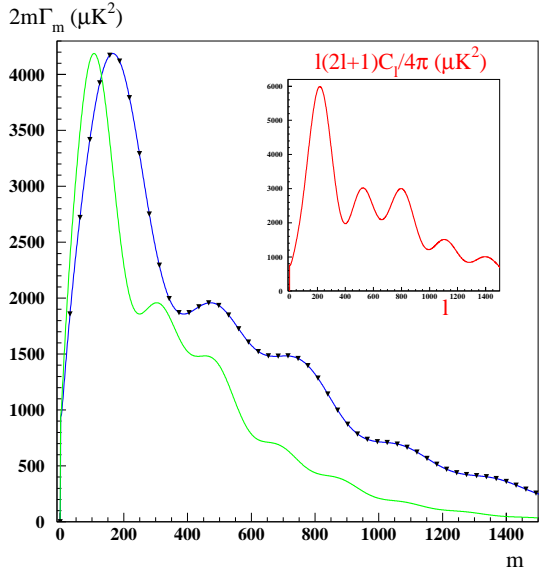,width = 8.cm}
\caption{'Typical' power spectra. 
The insert shows a 
$\ell (2\ell +1)C_{\ell}/4\pi$ spectrum up to $\ell =1500$.  
The main graphs are two Fourier spectra ($2m\Gamma_m$) exactly calculated 
using Eq.~\ref{eq:defgm} : one for 
$\Theta = 90^{\circ}$ (darker curve), and the other for 
$\Theta = 40^{\circ}$ (lighter curve).  
The triangles represent a subsample
of the $2 m \Gamma_{m}(\Theta= 40 ^{\circ})$ coefficients after having
rescaled their abscissa  by a factor $1/\sin 40^{\circ} \ =\  1.556$.}
\label{fig:spectres}
\end{figure}

Fig.~\ref{fig:spectres} shows an example of the $C_{\ell}$
power spectrum for $\ell <1500$,
together with two Fourier spectra\footnote{Note  that we have 
chosen the following normalizations: the $C_{\ell}$ coefficients
have been multiplied by $\ell(2\ell+1)/4\pi$ and the $\Gamma_{m}$
by $2m$.} which
describe the same sky for two quite distinct cases, one for 
$\Theta= 90 ^{\circ}$ 
and one for $\Theta = 40 ^{\circ}$.

Note that for this Fig.~\ref{fig:spectres} and throughout the article
the $C_{0}$  and $C_{1}$ coefficients have been set equal to $0$.

The relation that gives the $\Gamma_m(\Theta)$ from the $C_{\ell}$ was
obtained by \citet{del}:
\begin{eqnarray}
\label{eq:defgm}
\Gamma_{m}(\Theta) = \sum_{\ell =|m|}^{\infty}C_{\ell}B_{\ell}^{2}
\mathcal{P}_{\ell m}^{2}(\cos\Theta)\,,
\end{eqnarray}
where the set of $B_{\ell}$ coefficients characterizes the beam 
function and the $\mathcal{P}_{\ell m}^{2}$ are the \emph{normalized}
 associated
Legendre's functions. 
This relation assumes that the $a_{\ell m}$ introduced
in Eq.~\ref{eq:one} are uncorrelated Gaussian random variables
and that the scan is performed with a symmetric beam. 

In this article, we present the scaling law
and the inverse transformation that consists in
the calculation of the $C_{\ell}$ from the $\Gamma_{m}$.  In section 2, we
demonstrate that this simple scaling law, 
displayed
by the $m\Gamma_{m}$ spectrum
for different colatitude angles,
is accurate.  Section 3 is dedicated to
the description of two different methods 
proposed to invert Eq.~\ref{eq:defgm}  in the case 
$\Theta=90^{\circ}$.  While a simple matrix
inversion leads to the result, we also present an approximate analytic
method. In section 4  these two
methods are extended to the general case where $\Theta<90^{\circ}$.

\section{ Scaling of the $\bmath{\lowercase{m}\Gamma_{\lowercase{m}}(\Theta)}$ spectrum}\label{sec:scaling}


Our study was triggered by one of us noticing 
that the product $m\Gamma_{m}(\Theta)$ 
is only function of the reduced variable
$\mu \equiv {m}/{\sin \Theta}$, i.e. this product
 is independent (to a very
good approximation) of the colatitude angle
$\Theta$.

\begin{figure}
\epsfig{file=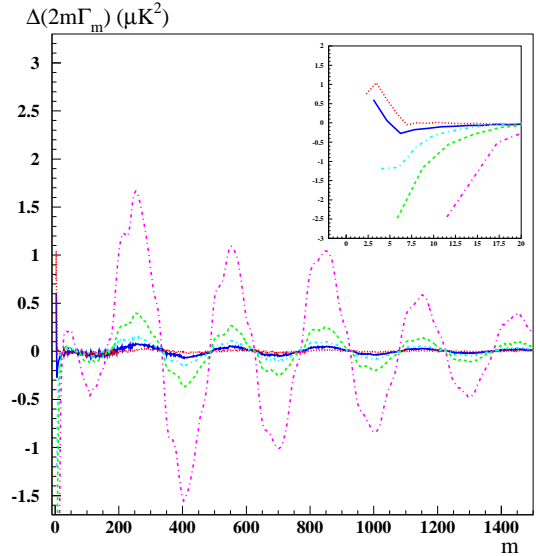,width = 8.cm}
\caption{ 
Absolute differences (in $\mu K^2$) between $2m\Gamma_m(\Theta)$ 
spectra scaled to $\Theta\ =\ 90^\circ$ and the interpolated 
 $2m\Gamma_m(\Theta=90^\circ )$  spectrum. All spectra are based on the $C_{\ell}$ 
spectrum of Fig.\ref{fig:spectres}. We have worked out these differences for 
$\Theta\ =\ 60^\circ$ (smallest amplitude curve), 
$40^\circ$, $30^\circ$, $20^\circ$ 
and $10^\circ$ (largest amplitude curve). 
The insert displays the low $m$ part, showing that 
the difference has a meaningful value only above $m=2/\sin \Theta$. 
}
\label{fig:spectresbis}
\end{figure}

This scaling is illustrated in Fig. \ref{fig:spectres} 
where a $2m\Gamma_m$ spectrum computed for a 
colatitude angle of $\Theta\ = \ 40 ^\circ$ is scaled to match the 
corresponding $\Theta\ = \ 90 ^\circ$ one. 
To quantify the precision of this approximate scaling law, 
we have computed the differences between the scaled $2m\Gamma_m(\Theta)$ 
and the interpolated $2m\Gamma_m(\Theta=90^\circ)$ spectrum 
(at $m/\sin \Theta$). Examples are shown in Fig. \ref{fig:spectresbis} 
for five $\Theta$ values ranging between $60$ and $10^\circ$. 
The absolute values of these 
differences are lower than $2\ \mu K^2$ over the whole $m$ range 
for the particular spectrum given
in figure \ref{fig:spectres}.
They are only defined for $m$ values greater than $2/\sin \Theta$, 
as shown in the insert. Oscillations
are observed in the difference. 
They present the same period but their amplitudes increase
as the colatitude angle $\Theta$ decreases.

Different $2m\Gamma_m(\Theta)$ 
sets obtained from several detectors over a small range of colatitude 
angles $\Theta$  (a few degrees)
may be combined using this scaling law, with a precision better than 
0.01\% . 
Several experiments, spanning a wider range of colatitude angles, 
may also be combined likewise, however with a slightly worse precision. 

In the following, 
we explain this scaling law
using a geometrical and a mathematical argument. 

\subsection{Geometric interpretation}

\noindent

The power spectrum  $\Gamma_{m}(\Theta)$ is the Fourier transform of the
signal autocorrelation function $A(\delta\phi,\Theta)$, where $\delta\phi$
 is the phase difference between two points of
the scanned ring. Two such points 
have an angular separation $\delta\psi$ on the
unit sphere, where:
\begin{eqnarray}
  \delta\psi=2 \arcsin(\sin\Theta\ \sin\frac{\delta\phi}{2})\,.
\end{eqnarray}

This relation between $\delta\phi$ and $\delta\psi$ allows one to express 
the scaling law, since the signal autocorrelation function,
expressed as a function of $\delta\psi$ is equal to the autocorrelation
function on a large circle scan:
\begin{eqnarray}
A(\delta\psi,\pi/2)= A(\delta\phi,\Theta)\,.
\end{eqnarray}

For small $\delta\phi$, this relation becomes linear:
\begin{eqnarray}
\label{eq:linear}
  \delta{\psi}=\sin\Theta\ \delta\phi\,.
\end{eqnarray}

So that, in this linear regime, the autocorrelation function satisfies:
\begin{eqnarray}
\label{eq:autocorr}
A(\delta\phi,\Theta)=A(\sin\Theta\ \delta\phi,\pi/2)\,.
\end{eqnarray}
Since the ring length $L$ on the unit sphere is $2\pi \sin \Theta$,
the $m^{\rm{th}}$ harmonic of the Fourier expansion
corresponds to 
structures on the sky of angular size 
\begin{eqnarray}
\label{eq:eqlambda}
\lambda \equiv 2\pi \frac{\sin \Theta}{m}= 
\frac{2\pi}{\mu}\,.
\end{eqnarray}

In the continuum approximation, 
taking the Fourier transform of both sides of 
Eq.~\ref{eq:autocorr} leads to:
\begin{eqnarray}
\label{eq:debscale}
\Gamma_{m}(\Theta)=\frac{1}{\sin\Theta}\ 
\Gamma_{m/\sin\Theta}(\pi/2)\,,
\end{eqnarray}
which, using Eq. ~\ref{eq:eqlambda} leads to the scaling law:
\begin{eqnarray}
\label{eq:scale}
m\Gamma_{m}(\Theta)=\mu\Gamma_{\mu}(\pi/2)\,.
\end{eqnarray}


While we are mainly concerned here 
 with circular scanning, the same reasoning
can be made for any kind of trajectory on the sky as long as it stays
`close' to a large circle on angular scales of order $\lambda$,
and the same scaling law applies to the power density spectrum 
expressed as a function of ${1/\lambda}$.

\subsection{Analytic interpretation}
To investigate this scaling mathematically, we start from Eq.~\ref{eq:defgm} which gives the 
exact relations that connect the $\Gamma_{m}(\Theta)$ to the
$C_{\ell}$. Since the $B_{\ell}$ are -- supposedly -- well known quantities
for each experimental set up, we will no longer mention them explicitly
 and we will deal with the coefficients \mbox{$\mathcal{C}_{\ell} 
\equiv
C_{\ell}B_{\ell}^{2}$.}\

We calculate the $\mathcal{P}_{\ell m}^{2}(\cos\Theta)$ factors 
using  approximate
expressions of the Legendre's associated functions given by \citet{rob} 
(see appendix A for some details) which, once normalized, read\

\noindent $\bullet$ for $\ell < m/\sin \Theta\,$:
\begin{eqnarray}
\nonumber
\mathcal{P}_{\ell m}&(\cos\Theta) &\simeq
 \frac{1}{2\pi}
\sqrt{\frac{\ell+\frac{1}{2}}{M}}\left( \frac{\ell\cos\Theta + M}{\ell} \right)
^{\ell+\frac{1}{2}}\\
\nonumber
&\times &\left( \frac{m\cos\Theta - M}
{(\ell-m)\sin\Theta} \right)^m\prod_{k=1}^m \sqrt{\frac{\ell+k-m}{\ell+k}} 
\,,\; \hspace{.05cm}(12\rm{a})
\end{eqnarray} 
\noindent where $M = \sqrt{m^2 - \ell^2 \sin^2 \Theta} $,\

\noindent  $\bullet$ and for $\ell >m/\sin \Theta\,$: \

\begin{eqnarray*}
\mathcal{P}_{\ell m}(\cos\Theta)  \simeq \frac{(-1)^m}{2\pi}
\sqrt{\frac{2(2\ell+1)}{N}} \cos \omega\, ,\hspace{2.05cm} (12b)
\end{eqnarray*}

\noindent where $N = \sqrt{l^2\sin^2\Theta-m^2}$. 
The expression of the
angle $\omega$ is given in appendix A. These approximations
are illustrated by Fig.~\ref{fig:plm}. 

For  $\ell < m/\sin \Theta$, the numerical value 
of $\mathcal{P}_{\ell m}^{2}(\cos\Theta)$ is negligible, while
for $\ell >m/\sin{\Theta}$\  Eq.\ 6b implies 
\setcounter{equation}{12}
\begin{eqnarray}
\mathcal{P}_{\ell m}^{2}(\cos\Theta)  
\simeq \frac{1}{4\pi^{2}}
\frac{2\ell + 1}{(\ell^{2} \sin^{2}\Theta - m^{2})^{1/2}}
[1 + \cos (2 \omega)] \,.
\end{eqnarray}

Since the CMB angular power spectrum varies slowly as
a function of $\ell$, we may replace the sum over $\ell$
in Eq.~\ref{eq:defgm} by an integral. We thus obtain
\begin{eqnarray}
\label{eq:gmapprox}
m\Gamma_{m}(\Theta) = \frac{m}{4\pi^{2}} 
\int_{\frac{m}{sin \Theta}}^{\ell _{max}}
\!
\frac{\mathcal{C}(\ell)[2\ell + 1][1 + \cos (2\omega)]}{(\ell^{2} \sin^{2}\Theta - m^{2})^{1/2}}\
 \, d\ell \,
\end{eqnarray} 
where  $\ell_{max}$ is an $\ell$ value beyond which the power
spectrum vanishes, and
$\mathcal{C}(\ell)$ is a function of $\ell \,\in \:]0,\ell_{max}]$
that smoothly interpolates the $\mathcal{C}_{\ell}$ coefficients
(a simple way of proceeding is given in Appendix~B).

\begin{figure}
\epsfig{file=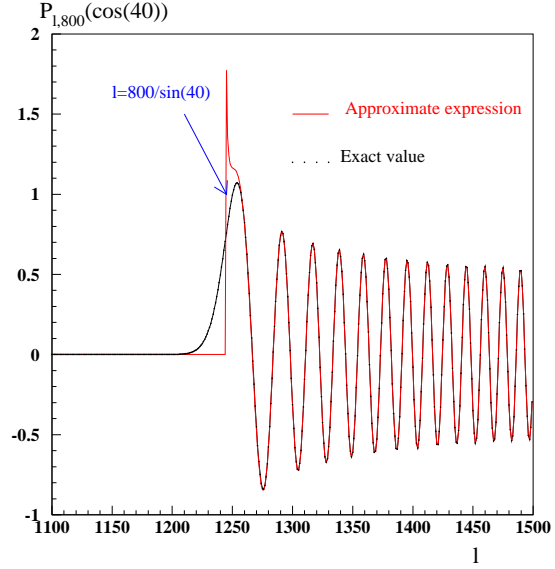,width = 8.cm}
\caption[]{Comparison between the exact value of 
$\mathcal{P}_{\ell\: 800}(\cos (40^{\circ}))$ 
as a function of $\ell$, dotted line, and the one 
obtained with the approximate expressions of
Eqs.~(12a) and (12b), solid line. The arrow indicates 
the $\ell\ = 800/\sin{40^\circ}$ abscissa.
}\label{fig:plm}
\end{figure}
The oscillation frequency $\nu$ of the cosine term (as a function of
$\ell$) in the integrand in the
right side of
Eq.~\ref{eq:gmapprox} is of order $\Theta/\pi$ (thus $\nu \sim 1/2$ when
$\Theta = \pi /2$). Such a
frequency is high enough for this cosine term to
contribute to a very small amount to the integral. This will be checked
numerically in section~\ref{sec:check}  below. Thus we may write: 
\begin{eqnarray}
\label{eq:gmintg}
m\Gamma_{m}(\Theta) \simeq 
\frac{1}{4\pi^{2}} \int_{\mu}^{\ell_{max}}\mathcal{C}(\ell)
\frac{2\ell + 1}{[(\ell/\mu)^{2} - 1]^{1/2}}\, d \ell\;. 
\end{eqnarray}
This equation demonstrates -- within the approximations 
that have been made -- that the product $m\Gamma_{m}(\Theta)$
depends only on the variable $\mu = {m}/{\sin \Theta}$.

Since the variable $\mu$ is not constrained to be an integer,
one has to introduce a smooth
function, $\Gamma (m, \Theta)$ where $m$ is now a real, that 
interpolates the $\Gamma_{m}(\Theta)$
discrete spectrum.
This can be done in the same way as the one indicated for the
$\mathcal{C}_{\ell}$ spectrum (cf. Appendix~B). 

 In terms of this  $\Gamma (m, \Theta)$ function, the 
scaling law is expressed by the relation:
\begin{eqnarray}
\label{eq:scaling}
\Gamma (m^{\prime}, \Theta^{\prime}) = \frac{\sin \Theta}{\sin \Theta^{\prime}}
\Gamma(m^{\prime} \frac{\sin \Theta}{\sin \Theta^{\prime}},\Theta)\,.
\end{eqnarray}
This equation follows from the equality $m \Gamma(m,\Theta) =
m^{\prime}\Gamma(m^{\prime}, \Theta^{\prime})$ which holds true provided that
$m/\sin \Theta = m^{\prime}/\sin \Theta^{\prime}$.

Assuming that the Fourier spectrum has been obtained for a particular
value $\Theta$ of the colatitude angle, Eq.~\ref{eq:scaling} allows one to
calculate $\Gamma(m^{\prime}, \Theta^{\prime})$ for 
$m' = m \sin \Theta^{\prime}/\sin \Theta , \;
m = 1,2 \cdots  m_{max}=\ell_{max}\sin \Theta$. Then, by interpolation, 
one gets $\Gamma(m^{\prime},\Theta^{\prime})$  for
all integer values of $m^{\prime}$ ranging from $\sin \Theta ^{\prime}/
\sin \Theta$ up to $\ell_{max}\sin \Theta^{\prime}$. Eq.~\ref{eq:scaling}
 can thus
be used to compare and combine Fourier spectra that correspond to 
different $\Theta$ values. 

\section{Recovering  the $\bmath{\mathcal{C}_{\ell}}$ coefficients
from the $\bmath{\Gamma_{\lowercase{m}}(\pi/2)}$ Fourier spectrum}

\subsection{Checking and solving the integral equation that relates 
$\bmath{\mathcal{C}(\ell)}$ to $\bmath{\Gamma (\lowercase{m}, \pi/2)}$ \label{sec:check}}

\noindent
Since $\Theta$ is assumed to be equal to $\pi/2$ in this section, the
variable $\mu$ can be identified with $m$.

In order to facilitate the numerical calculation
of the right side of Eq.~\ref{eq:gmintg}, we
introduce a new variable of integration $x$ defined by
$\ell = m \cosh x$. Then Eq.~\ref{eq:gmintg} can be rewritten  
\begin{equation}
\label{eq:gmtransf}
\Gamma(m,\pi/2) = \frac{1}{4 \pi^{2}}\int_{0}^{\cosh^{-1} (\ell_{max}/m)} 
\hspace{-1.4 cm}(2m\cosh x+1)\mathcal{C}
(m\cosh x)\; dx\,.
\end{equation}

\begin{figure}
\epsfig{file=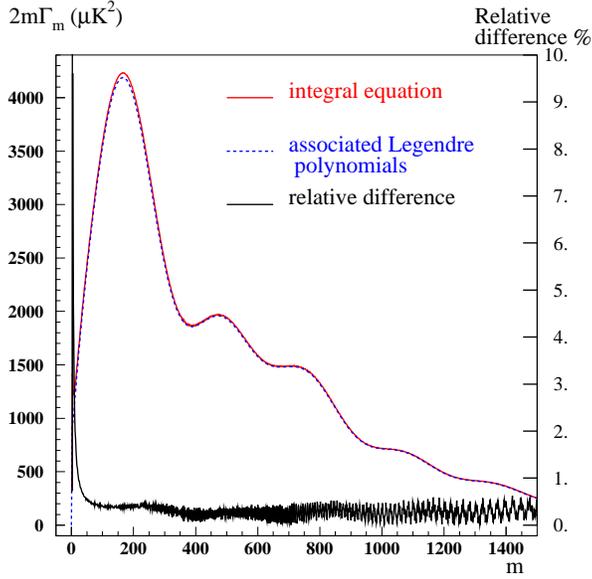,width = 8.cm}
\caption{ Comparison 
between the $2 m \Gamma_{m}$ coefficients computed with the 
associated Legendre's polynomials (dashed line)
and the $2 m \Gamma(m)$ function calculated using Eqs.~\ref{eq:gmtransf}, 
\ref{eq:clinterp} and \ref{eq:interpf} with $\sigma = 0.5$ (solid line). 
The relative 
difference between the results of the two calculations is 
shown by the lower curve (in \%, right scale).}\label{fig:cltogam}
\end{figure}

The transformation defined by Eq.~\ref{eq:gmtransf} is 
linear: thus one may insert in the integrand an interpolating function 
of the $\mathcal{C}_{\ell}$ spectrum as defined by Eq.~\ref{eq:clinterp}.
The output of 
Eq.~\ref{eq:gmtransf} applied to the angular power spectrum
of Fig.~\ref{fig:spectres} is shown in
 Fig.~\ref{fig:cltogam}.  One can see in this Fig.~\ref{fig:cltogam}
that for such a spectrum the 
approximations that were made in section~\ref{sec:scaling}
ensure an accuracy better than
1\% --  except at the lower end of the
spectrum where the relative error drops
below $2$ \% for $m = 14$.

Eq.~\ref{eq:gmtransf} can be solved for $\mathcal{C}(\ell)$
by noticing that
this integral equation is similar to Schl\"{o}milch's equation
 which reads
\begin{eqnarray}
F(m) = \frac{2}{\pi} \int_{0}^{\pi/2} \Phi (m \sin x)\; dx
\end{eqnarray}
\noindent where $m$ is a real.

The way to solve the latter equation can be found {\it e.g.} in
\citet{kra}. We proceed in a similar way  for Eq.~\ref{eq:gmtransf}  
(the details are given in Appendix~C) and we obtain 
\begin{eqnarray}
\label{eq:clreslt}
{\mathcal {C}({\ell}) = - 8\pi \frac{\ell}{2\ell+1}
\int_{0}^{\cosh^{-1} (\ell_{max}/\ell)}  \Gamma ^{\prime} 
(\ell \cosh x)\; dx}\,,
\end{eqnarray}

\noindent where $\Gamma ^{\prime}$ is the derivative of $\Gamma (m,\pi/2)$ 
with respect
to $m$.
Again the transformation implied by Eq.~\ref{eq:clreslt} is a linear one, allowing
the use of interpolating functions as defined in Appendix B.
Fig.~\ref{fig:gamtocl} illustrates the use of this integral equation
to calculate the $\mathcal{C}_{\ell}$ coefficients starting 
with the set of $\Gamma_{m}(\pi/2)$'s.\\
\begin{figure}
\epsfig{file=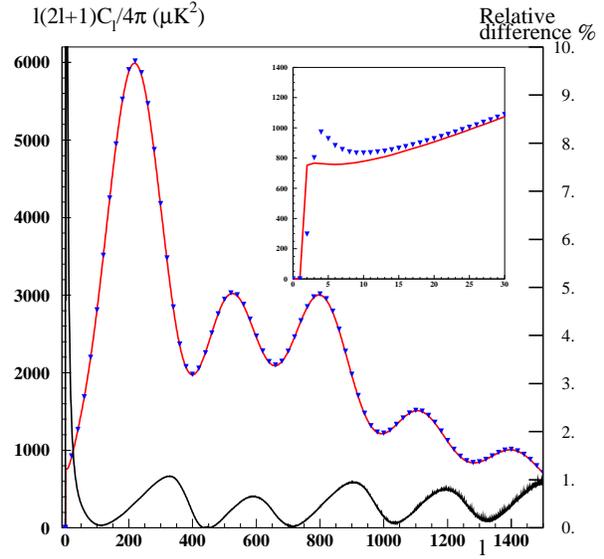,width = 8.cm}
\caption[]{ Comparison
between the 'Typical' $\mathcal{C}_{\ell}$ coefficients 
for $\Theta=90^\circ$ (solid line), 
used to calculate the 
$\Gamma_m$ Fourier spectrum
(using the \Pmat matrix)
and the $\mathcal{C}(\ell)$ function obtained by inserting 
this Fourier spectrum  
in Eq.~\ref{eq:clreslt}, (triangles; only some points are shown). We have 
set $\sigma = 1$ in Eq.~\ref{eq:interpf}. The relative
difference (in \%) is shown by the lower curve (right scale).
Insert: zoom on the low $\ell$ region.}
\label{fig:gamtocl}
\end{figure}

\subsection{Numerical inversion \label{sec:pmat}}

In the $\Theta={\pi}/{2}$ case, the connection between the set of
\cl's and the corresponding $\Gamma_{m}$'s is simple since
Eq.~\ref{eq:defgm} can be written using matrices (\citealt{Piat}):
\begin{eqnarray}
  \vec{\Gamma}= \Pmat \times \vec{\mathcal{C}}\,,
\end{eqnarray}
with:
\begin{eqnarray}
  \Pmat_{ij}=\left[{\cal P}_{ji}(0)\right]^2,
\end{eqnarray}
where ${\cal P}_{ji}$ are the normalized associated 
Legendre's functions.
\Pmat  is (upper) triangular. 

In addition, since
the associated Legendre polynomials are defined as:


\begin{eqnarray}
P_{\ell m}(0)=
\begin{cases}
(-1)^{p} \frac{(2\ell +2m)!}{2^{\ell}p!(p+m)!} & \mbox{if\ $\ell-m=2p$}\hspace{1cm} \mbox{[a]} , \\
0 & \mbox{if\ $\ell-m=2p+1$}\hspace{.45cm}  \mbox{[b]},
\end{cases}
\end{eqnarray}


\noindent all of the $\Pmat_{ii}$ diagonal elements are different from zero --
thus this matrix is invertible.

The inverse of \Pmat is also upper triangular and keeps the
peculiar structure of the original matrix: in both \Pmat and
\mbox{\Pmat\!\!$^{-1}$}
only the $\ell-m=2p$ terms differ from zero. 

\subsection{Comparison between the analytic and 
the numerical transformations}

\begin{figure}
\epsfig{file=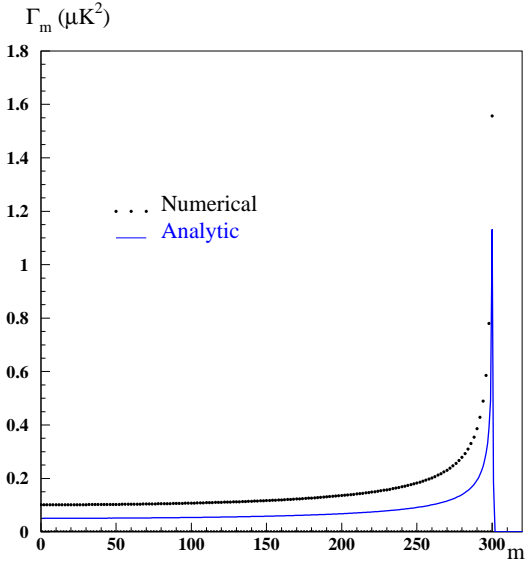,width = 8.cm}
\caption[]{ Middle curve (solid line) : Fourier spectrum
obtained using Eq.~\ref{eq:gmtransf} when only 
$\mathcal{C}_{300}\neq 0$. We have used $\sigma = 0.5$ in 
Eq.~\ref{eq:interpf}. Upper and 
lower set of points: the $\Gamma_m$ coefficients computed
with the $\Pmat $ matrix. 
Since we assume here that
$\Theta = \pi/2$, all $\Gamma_{m}$ coefficients
whose indexes $m$ are odd vanish.}
\label{fig:greencl}
\end{figure}

 One way of comparing
the two methods of calculating
the Fourier spectrum is to look at what happens 
when a single $\mathcal{C}_{\ell}$  coefficient
is different from zero. This is done in
 Fig.~\ref{fig:greencl} for the case where $\mathcal{C}_{300}=1$.
Note that since we assume here that
$\Theta = {\pi}/{2}$,  Eq.~22\ b implies
that all $\Gamma_{m}$ coefficients with an odd index vanish 
(for a single non vanishing $C_{\ell}$ coefficient with an odd
$\ell$ value, all $\Gamma_{m}$ coefficients
with an even index would vanish).
One notices that the $\Gamma (m)$
function  runs at mid-height of the non-vanishing
$\Gamma_{m}$ coefficients.\

\begin{figure}
\epsfig{file=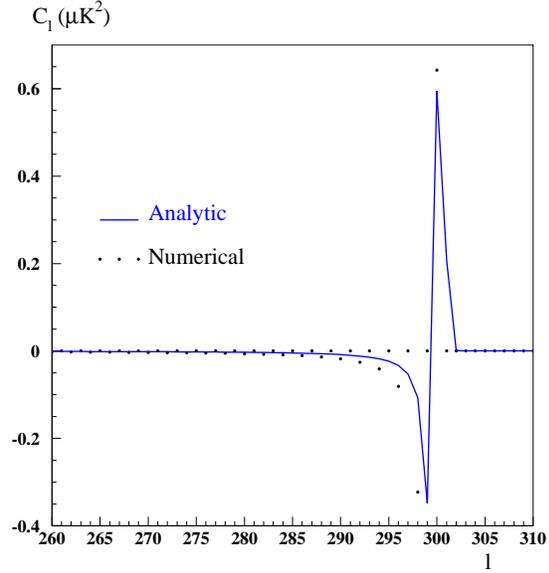,width = 8.cm}
\caption[]{Solid line: the $\mathcal{C}_{\ell}$ spectrum
obtained with Eq.~\ref{eq:clreslt} when only 
$\Gamma_{300}\neq 0$  (we have set 
$\sigma = 1$ in Eq.~\ref{eq:interpf}). In dots: the 
$C_{\ell}$ coefficients
calculated with the \Pmat\!\!$^{-1}$ matrix . 
 }
\label{fig:greengm}
\end{figure}

Conversely, one may look at the $\mathcal{C}(\ell)$ function
that corresponds to the case where a single $\Gamma_{m}$ Fourier coefficient
is different from zero as shown in  Fig.~\ref{fig:greengm}
(here we used $\Gamma_{300}\neq 0$). The fact that
the $\mathcal{C}(\ell)$ graph is negative in some domain of $\ell$ values
shows that no distribution of temperature inhomogeneities which
satisfies the validity conditions of Eq.~\ref{eq:defgm}
(isotropy and Gaussian $a_{\ell m}$) can correspond to a Fourier
spectrum with a single non-vanishing coefficient.\

Taken together, Figs.~\ref{fig:greencl} and \ref{fig:greengm} show where 
we should expect a 
strong signal in one spectrum in the case the other spectrum 
presents
a high power in some particular bins.\\


\section{Working with smaller rings on the 
sky ($\bmath{\Theta < \pi/2}$)}\label{sec:small}

\subsection{General features of the Fourier spectrum}\label{sec:genfea}

 In the preceding section we assumed that the scanned rings 
are the largest ones on the sphere
($\Theta = {\pi}/{2}$). In this
case the fact that the \Pmat matrix is invertible 
 establishes that the
Fourier spectrum of such rings contains all the physical information carried
by the $\mathcal{C}_{\ell}$ coefficients.\

Scanning smaller circles on the sky implies a higher fundamental frequency in 
Fourier angular space and thus a less dense sampling of this 
Fourier space. \

In fact\vspace{-3mm}
the loss of information is then twofold:
\begin{itemize}
\item The  $G(\mu)\equiv m\Gamma (m,\Theta)$ function is no longer
measured for $\mu =1$: the lowest value of $\mu$ which can be reached with
the data is now $\mu = {1}/{\sin \Theta}$. \

\item Secondly, $G(\mu)$ is no more measured
for $\mu$ values
that differ by one but for $\mu$ values that
differ by ${1}/{\sin \Theta}$. As a very simple example:
if the scan is performed for $\Theta = {\pi}/{6}$, then one
measures $G(\mu)$ only for $\mu= 2n$ with $n\, \in\, ]0,\ell_{max}/2] $.
Because of  
the smoothness of the angular spectra, this sparse sampling of 
the function $G(\mu)$ is not necessarily a drawback as long 
as the accuracy of the measurements compensates for it.  
\end{itemize}

\subsection{Analytic calculation of the $\bmath{\mathcal{C}_\ell}$ 
spectrum for $\bmath{\ell > 1/\sin \Theta}$} 

As far as the analytic calculation of the 
$\mathcal{C}_{\ell}$ spectrum
is concerned, it can be performed with the same formalism as above (cf.
subsection~\ref{sec:check}).
One should merely replace the derivative of $\Gamma(m,\pi/2)$
that appears in the right side of Eq.~\ref{eq:clreslt} by
the derivative (with respect to $m$) of 
\setcounter{equation}{22}
\begin{eqnarray}
\tilde{\Gamma}(m)= \sin \Theta \sum_{i = 1}^{\ell_{max}\sin \Theta} \Gamma_{i}
f(m\sin \Theta- i)\,.
\end{eqnarray}
\noindent $\tilde{\Gamma}(m)$ is just the rescaled 
version (cf. Eq.~\ref{eq:scaling}) of $\Gamma (m, \Theta)$ 
defined by  Eq.~\ref{eq:apndxb}
(this rescaling translates the $\Gamma (m,\Theta)$ Fourier spectrum
 into the one corresponding to $\Theta = \pi/2$).
Furthermore the width of the interpolating function $f(x)$ of Appendix B
(see Eq.~\ref{eq:interpf}) should be
increased by a factor ${1}/{\sin \Theta}.$\\

\subsection{Numerical calculation of the $\mathcal{C}_\ell$ spectrum 
for $\ell > 1/\sin \Theta$}

It follows from subsection~\ref{sec:genfea}
above  that the $\Gamma_{m}(\Theta)$ coefficients
differ significantly from zero in the range
$1\le m \le \ell_{max}\sin\Theta$. 
Then using the $\Gamma (m,\Theta)$ 
function that interpolates these coefficients and Eq.~\ref{eq:scaling} 
one can calculate
the following set of \mbox{$\ell_{max}-\ell_{min} +1$} values 
\begin{eqnarray}
\label{eq:gamtilde}
  \tilde{\Gamma}_{m^\prime}=\sin\Theta\ \G{m^\prime\sin\Theta,\Theta},
\end{eqnarray}
with $m^\prime= \ell_{min},\, \ell_{min}+1,\cdots\ell_{max}$
where $\ell_{min}$
is the first integer larger than $1/\sin \Theta$. 
These $\tilde{\Gamma}_{m^\prime}$ coefficients are the ones
of the Fourier spectrum for
$\Theta = \pi/2$. Once obtained, the \cl spectrum
is simply given by
\begin{eqnarray}
\label{eq:inverse}
  \vec{\mathcal{C}}= \Pmat^{-1}\vec{\tilde{\Gamma}}\,
\end{eqnarray}
for $\ell \geq \ell_{min}$.
The \Pmat matrix and its inverse have been discussed in section
\ref{sec:pmat}. The first $\ell_{min}-1$
rows and columns of $\Pmat^{-1}$ should be omitted in Eq.~\ref{eq:inverse}
since the lowest value of the  $m^\prime$ index is $\ell_{min}$.

\begin{figure}
\epsfig{file=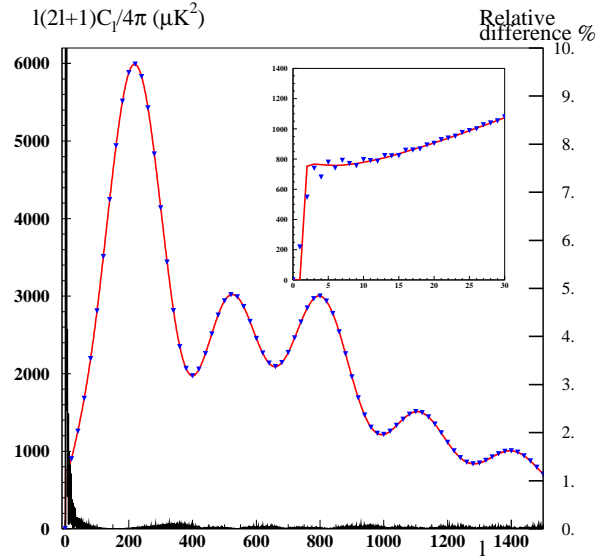,width = 8.cm}
\caption[]{ The input \cl spectrum (solid curve) and the one
  reconstructed by the numerical method in the $\Theta=40^\circ$
  case (only some points are shown). The relative difference between the two
spectra is shown by the lower curve (in \%, right scale). Insert : zoom
on the low $\ell$ region.}
\label{fig:num40}
\end{figure}

Fig.~\ref{fig:num40} 
shows a numerical example: we use the `typical' \cl spectrum
of Fig.~\ref{fig:spectres}, to produce a set of $\Gamma_{m}$ values in the 
$\Theta=40^\circ$ case (Eq.~\ref{eq:defgm}). 
Then we apply the method described above and
compare the input spectrum to the obtained one. In this 
example we used a simple linear interpolation of the
$\Gamma_{m}$ spectrum. The agreement is excellent and 
better than the one obtained with the analytic
method (cf. Fig.~\ref{fig:gamtocl}) as the latter
involves some approximations (cf. section 2) in addition to the
ones stemming from the scaling and the interpolation procedure.

The excellent agreement of Fig. \ref{fig:num40} 
breaks down for low values of
$\ell$. Nevertheless, for $\ell\gtrsim 3$ in the $\Theta=40^\circ$ case,
one gets an agreement better than 10\% (far above the
cosmic variance).
For the case $\Theta=80^\circ$ our simple scaling method can
be used up to an accuracy better than 1\% for any $\ell$ values.


\section{Conclusion }

We have shown how data taken on 
circles with different colatitude angle
$\Theta$ can be combined using a scaling law that is
satisfied by the 
$m\Gamma_{m}(\Theta)$ coefficients
at the  $0.1$ \% level in a wide range of $m$ and $\Theta$ values.\

Then we have derived this scaling property from  both
geometrical considerations and 
linear expressions
of the $\Gamma_{m}$ coefficients in terms of the $C_{\ell}$ ones
by introducing  analytic approximations of the normalized
Legendre's associated polynomials $\mathcal{P}_{\ell m}(\cos\Theta)$
that enter these relations.\

Integral equations were obtained that 
relate to a good
approximation  interpolating
functions of the two sets of coefficients ($\Gamma_{m}$ and
$C_{\ell}$). These analytic relations give a simple picture 
of the connection between the two types of spectra
and are easy to use.

Finally  we have investigated ways
of calculating the $C_{\ell}$ coefficients  when the  $\Gamma_{m}$ Fourier 
spectrum is known.
We have shown how the 
inverse of the $\mathcal{P}_{\ell m}^{2}(0)$ matrix can be used
 to perform this calculation not
only for $\Theta = \pi/2$ but also in the general case
where $\Theta < \pi /2$. This was achieved by taking advantage
of the scaling of the $m\Gamma_{m}$ spectrum on the one hand
and of its smoothness on the other. 

This set of results provides a  basis for further
investigation of the connection between the 
\hspace{-1mm}\emph{ measured} $C_{\ell}$
and $\Gamma_{m}$ spectra altered by noise and errors.\\

\appendix
\section{Approximate expressions of the normalized Legendre's
associated polynomials} 

We start with asymptotic expressions
of the 
Legendre's functions obtained  by ~\citet{rob} in the limit
of large $\ell$, ${m}/{\ell}$ being kept
constant. These asymptotic expressions depend on the relative value of $m$
and $\ell \sin \Theta$.\

\noindent $\bullet$
For $\ell <m/\sin \Theta$,\
\begin{eqnarray}
\nonumber
P_{\ell m}(\cos \Theta) \simeq &    \\
\frac{(-1)^m\ell!}{\sqrt{2\pi}(\ell-m)!}&\frac{(\ell\cos \Theta+M)
^{\ell+\frac{1}{2}}(m\cos \Theta - M)^m}{\ell^{\ell+
\frac{1}{2}}(\ell-m)^mM^{\frac{1}{2}}\sin^m \Theta}\,,   
\end{eqnarray}
where  $M = \sqrt{m^2 - l^2 \sin^2 \Theta}\,, $\

\noindent $\bullet$  while for $\ell >m/\sin \Theta$,\
\begin{eqnarray}
\nonumber
P_{\ell m}(\cos \Theta)\simeq& \\
(-1)^m \sqrt{\frac{2}{\pi}}&\frac{\ell!(\ell-m)^{\frac{\ell-m}{2}+
\frac{1}{4}}(\ell+m)^{\frac{\ell+m}{2}+\frac{1}{4}}}
{(\ell-m)!\ell^{\ell+\frac{1}{2}}N^{\frac{1}{2}}} \cos \omega
\end{eqnarray}
where 
\begin{eqnarray}
N & = & \sqrt{\ell^2\sin^2\Theta-m^2}\,,\\
\omega & = & (\ell+\frac{1}{2})\alpha -m\beta-\frac{\pi}{4}\,,\\
\alpha & =  & arg(\ell\cos\Theta + iN)\,,\\
\beta & = & arg(m\cos\Theta + iN)\,.
\end{eqnarray}

To `normalize' these polynomials and obtain
the $\mathcal{P}_{\ell m}$ ones, they must be multiplied
by 
\begin{eqnarray}
\sqrt{\frac{2\ell+1}{4\pi}\frac{(\ell -m)!}{(\ell+m)!}}.
\end{eqnarray}

Then
the last step consists in using  Stirling's formula  
($n! \simeq \sqrt{2\pi} \,n ^{n+\frac{1}{2}}\,e^{-n}$) to
replace the factorials by analytic functions. A few simplifications
can then be made  that lead
to the approximate expressions used in section 2.\\
 
\section{Interpolating functions of the discrete power spectra \label{sec:interpol}}
\noindent 
Since the calculation of the $C_{\ell}$ coefficients
involves integrals over spherical Bessel $j_{\ell}$
functions (see, {\it e.g.} \citealt{SandZ}), one may 
try and use an expression of
these functions that extends them to non-integer 
values of $j$. But here we will adopt a much simpler
procedure and write 
\begin{eqnarray}
\label{eq:clinterp}
\mathcal{C}(\ell) \equiv \sum_{i = 1}^{\ell_{max}} \mathcal{C}_{i}
f(\ell-i),
\end{eqnarray}
where $\ell$ is now a real whose value ranges between $2$ 
(recall that we ignore
the dipole term) and $\ell_{max}$,
and $f(x)$ is a positive, infinitely differentiable function
($f\in\,{\cal C}^{\infty}$), 
which differs significantly
from $0$ in an $|x|$ range which is of order unity, and
whose integral over $x$ is unity. In practice
we used 
\begin{eqnarray}
\label{eq:interpf}
f(x) = \frac{1}{\sqrt{2\pi}\,\sigma}\exp {\frac{-x^{2}}{2\sigma^{2}}}
\end{eqnarray}
with $\sigma \sim 1$.\

Similarly, we define an interpolating function for
 the $\Gamma_{m}(\Theta)$ coefficients in the following way:
\begin{eqnarray}
\label{eq:apndxb}
\Gamma(m, \Theta)\equiv \sum_{i = 1}^{\ell_{max}\sin \Theta} \Gamma_{i}
f(m-i),
\end{eqnarray}
where $m$ is  a real and $f(x)$ is chosen
as above.

\section{Inversion of the integral equation relating
$\bmath{\mathcal{C}(\ell)}$ to $\bmath{\Gamma (\lowercase{m})}$\label{sec:invint}}

Since $\mathcal{C}(\ell)$ vanishes 
for $\ell > \ell_{max}$,  the integral equation (\ref{eq:gmtransf})
is of the form 
\begin{eqnarray}
\Gamma(m) = \int_{0}^{\infty}h(m\cosh x)\; dx\,.
\end{eqnarray}
We differentiate both sides of this equation with respect to $m$,
substitute for this variable $m$ the product $u \cosh \psi$, 
and integrate both sides over $\psi$ between the limits
$0$ and $\infty$. 
We thus obtain: 
 \begin{eqnarray}
\int_{0}^{\infty}\!\!\!\!\!\!\Gamma '(u \cosh \psi) \: d\psi =\!\!\!
\int_{0}^{\infty}\!\!\! \! \!\!dx \!\int_{0}^{\infty\!}\!\!\!
\!\!h'(u\cosh \psi \cosh x)
\cosh x \: d\psi.
\end{eqnarray}

Then  a new integration variable $\xi$ is used
in the second integral of the right side of this equation, 
defined by
$\cosh \xi =  \cosh \psi \cosh x$. Some simple algebra then leads to 
\begin{eqnarray}
\int_{0}^{\infty}\!\!\!\!\!\!\Gamma '(u \cosh \psi)  d\psi =\!\!\!
\int_{0}^{\infty}\!\!\!\!\!\! dx \int_{x}^{\infty}
\!\! \frac{h'(u\cosh \xi) \sinh \xi \cosh x}{\sqrt {\sinh^{2}\xi - 
\sinh^{2}x}} d\xi .
\end{eqnarray}
Once  the integration order is reversed in the right side of this equation
on obtains:
\begin{eqnarray} 
\nonumber
\int_{0}^{\infty}\Gamma '(u \cosh \psi) \; d\psi =
\int_{0}^{\infty} h'(u\cosh \xi) \sinh \xi d\xi \\
\int_{0}^{\xi}\frac{\cosh x \; dx}{\sqrt {\sinh^{2}\xi - \sinh^{2}x}}.
\end{eqnarray}
The integral over $x$ is simply ${\pi}/{2}$.
Furthermore  $h(\infty)= 0$ in our case, so that 
\begin{eqnarray}
h(u) = -\frac{2 u}{\pi} \int_{0}^{\infty}\Gamma ' (\xi \cosh \psi)\; d\psi\,.
\end{eqnarray}

\end{document}